
%
\magnification\magstep1
\hsize=14.5truecm
\vsize=23truecm
\hoffset=1.0truecm
\voffset=-1.0truecm
\baselineskip=0.5truecm
\parindent 8pt
\footline={\hss\tenrm-- \folio\ --\hss}	
\parskip 0pt 
%
%
\font\bigbf=cmbx10 scaled \magstep1

 at 11truept 
\font\it=cmti10 at 11truept 
\font\bf=cmbx10 at 11truept 
%
%

\def\hi{\noindent \hangindent=2.5em}
\def\etal{{\it et\thinspace al.}\ }    
\def\hmpc{\; h^{-1}\;{\rm Mpc}}

\def\lb{\hfill\break}
\def\ref{\par\noindent\hangindent 20pt}
\def\deg{\ifmmode^\circ\else$^\circ$\fi}    


\def\arcs{\ifmmode {'' }\else $'' $\fi}     
\def\arcm{\ifmmode {' }\else $' $\fi}     
\def\buildrel#1\over#2{\mathrel{\mathop{\null#2}\limits^{#1}}}
\def\mper{\ifmmode \buildrel m\over . \else $\buildrel m\over .$\fi}
\def\hper{\ifmmode \rlap.^{h}\else $\rlap{.}^h$\fi}
\def\sper{\ifmmode \rlap.^{s}\else $\rlap{.}^s$\fi}
\def\arcsper{\ifmmode \rlap.{' }\else $\rlap{.}' $\fi}
\def\arcmper{\ifmmode \rlap.{'' }\else $\rlap{.}'' $\fi}

%
%
\def\lb{\hfil\break}

\def\nat{{\it Nature }}
\def\apj{{ApJ}}

\def\mnras{{MNRAS}}

%
%
\def\2acapo{}               
\def\deg {^\circ}                       
\def\etal {\it et al. \rm}              

\def\nupa{\vfill\eject}        
\def\fig#1.#2{{\sl Figura~#1.#2}}       
\def\tab#1.#2{{\sl Tabella~#1.#2}}      
\def\lead{\leaders\hbox to 10pt{\hfill . \hfill}\hfill}


\def\deg{$^{\circ}$}

\def\<{$<$}
\def\>{$>$}

\def\H2{H$_2$}

\def\H0{H$_{\circ}$}
\def\tento#1;{$10^{#1}$}

\def\rref#1;#2;#3;#4;#5{#1, #2, #3, #4, #5}
\def\ref#1;#2;#3;#4{#1, {\it #2}, {\bf #3}, #4}
\def\book#1;#2;#3{\par\hangindent 20 pt #1, {\it #2}, #3}

\def\vvec{\hbox{\bf v}}
\def\lb{\hfill\break}


\null\vskip 1 truecm

{\bigbf
\centerline {MODELLING THE POWER SPECTRUM}
\centerline {OF DENSITY FLUCTUATIONS:}
\centerline {A PHENOMENOLOGICAL APPROACH}
}

\null\vskip 1 truecm

\centerline {Enzo Branchini$^{1}$, Luigi Guzzo$^2$ \&
Riccardo Valdarnini$^1$ }

\null\vskip 4.5 truecm
\centerline {Accepted by {\it The Astrophysical Journal (Letters)} --
December 30, 1993}

\null\vskip 4truecm

{\baselineskip 0.5truecm
\noindent $^1$ SISSA - International School of Advanced Studies,\lb
Strada Costiera 11, I-34014 Trieste, Italy.

\noindent $^2$ Osservatorio Astronomico di Brera, Sezione di Merate,\lb
Via Bianchi 46, I-22055 Merate (CO), Italy.

\phantom{.....}
}
\nupa

\centerline {\bf {ABSTRACT}}
\bigskip

We show how, based on considerations on the observed form of the galaxy
two--point spatial correlation function $\xi(r)$, a very simplified
-- yet surprisingly effective -- model for the linear density fluctuations
power spectrum can be constructed.  We first relate the observed
large--scale shape of $\xi(r)$ to a power--law form for the power spectrum,
$P(k)\propto k^{-2.2}$.   For a plausible value of the bias parameter
$b=1/\sigma_8\simeq1.8$, one has $(\delta \rho / \rho)_{rms}\sim 1$
at $r \simeq 3.5 \hmpc$, suggesting that the change of slope
observed in $\xi(r)$ around this scale
marks the transition between the linear and nonlinear gravitational
regimes.  Under this working hypothesis, we use a simple analytical
form to fit the large--scale correlations constraints together with the
COBE CMB anisotropy measurement, thus constructing a
simple phenomenological model for the linear power spectrum.
Despite its simplicity, the model fits remarkably well directly estimated
power spectra from different optical galaxy samples, and when evolved
through an N--body simulation it provides a good match to the observed
galaxy correlations.  One of the most interesting features of the model
is the small--scale one--dimensional velocity dispersion produced:
$\sigma_{1d}$ = 450 Km s$^{-1}$ at 0.5 $\hmpc$ and $\sigma_{1d}$ = 350
Km s$^{-1}$ for separations $\ge 2 \hmpc$.

\bigskip\bigskip

\noindent
{\bf 1. INTRODUCTION}
\smallskip

Cosmological models for the formation of the large--scale structure
of the Universe  require as a key ingredient a specific form for the
power spectrum of density fluctuations $P(k)$ at recombination.
This is usually specified in terms of a {\sl primordial spectrum} together
with a {\sl transfer function} which describes the subsequent evolution of
fluctuations depending on the physical scenario.
A paradigmatic example is provided by the Cold Dark Matter (CDM)
model (e.g. Blumenthal \etal 1984; White \etal 1987).

Conversely, one might ask, independently of any {\sl a priori} physical
model, whether a specific shape for the power spectrum is implied by
the available data, i.e. adopt a {\sl phenomenological} approach
(e.g. Kashlinsky 1992; Scaramella 1992; Taylor \& Rowan--Robinson 1992).
Indeed, observations have reached a stage in which it is possible to
constrain directly the power spectrum on linear scales, where
$(\delta \rho / \rho)_{rms}< 1$ and the recombination shape of
$P(k)$ should be preserved.   On one side
galaxy redshift surveys have allowed reliable estimates of the two--point
correlation function $\xi(r)$, the Fourier counterpart
of $P(k)$, up to separations $\sim 30 \hmpc$ (de Lapparent, Geller \&
Huchra 1988; Guzzo \etal 1991, hereafter G91; Loveday \etal 1992),
and direct estimates of $P(k)$ itself
(Baumgart \& Fry 1991; Peacock \& Nicholson 1991; Vogeley \etal 1992,
hereafter V92; Jing \& Valdarnini 1993, hereafter JV93; Fisher \etal 1993,
hereafter FDSYH).
On much larger scales,
the discovery by the COBE--DMR experiment of temperature fluctuations in the
microwave background radiation (Smoot \etal 1992) has produced a further
fundamental constraint on the very long--wavelength amplitude of $P(k)$.
Among the outcomes of this relative wealth of results is the fact that
the standard flat CDM model (although successful on several grounds),
cannot consistently reproduce all the observations.
The difficulties arise essentially from a too--high ratio of
small--scale to large--scale power.
Several modifications to the standard CDM scenario have been explored
in the attempt to overcome these problems .
Among these, open models with non--vanishing cosmological constant
(Efstathiou \etal 1990), models with {\sl tilted} primordial index
(Adams \etal 1992, hereafter A92; Cen \etal 1992), or mixed (CDM
plus HDM) models (Davis, Summers \& Schlegel 1992 and references
therein).
Rather than entering the debate on which physical model is closer to
the observed behaviour, in this {\it Letter} we prefer to keep to the pure
phenomenological approach to reconstruct a ``minimal'' model for the
linear $P(k)$.  As we shall see, when we close the circle, despite its
naivety the model proves to be remarkably effective in accounting for
a number of observations.

\bigskip\noindent
{\bf 2. A PHENOMENOLOGICAL POWER SPECTRUM}
\smallskip
The most natural way to construct a phenomenological model for the
linear $P(k)$ would be to start from its most recent direct
determinations (as e.g. V92 or JV93), try to `clean'
the observed shape from the expected non--linear effects, and
then add further constraints from other observables.
Here, however, we choose
to start in an apparently less straightforward
way, i.e. from the observed two--point correlation function $\xi(r)$.
The reason is that we first want to show how the large--scale
drop--off of $\xi(r)$ contains important information and,
although not a power--law, can be related to a specific,
simple shape of $P(k)$ on intermediate scales.
We stress that the phenomenological model obtained from this heuristic
procedure will be fully justified only {\it a posteriori}, when it will
possibly satisfy the main observational constraints.

Let us therefore consider the observed spatial correlation function
estimated from redshift survey samples.   This is different
from the true (real space) one, because on small scales large velocity
dispersions within clusters (`Fingers of God') depress its amplitude,
while on large scales coherent motions
amplify it according to linear theory prescriptions (Kaiser 1987).
These effects can be properly accounted for by estimating
$\xi(r_p,\pi)$, (e.g. Fisher \etal 1993b), i.e. decomposing the
pair separation vector along the directions parallel and perpendicular
to the line of sight.  A practical, approximate correction of the small--scale
distortions can however be performed by collapsing the cluster `fingers'
into a region corresponding to their statistically expected spatial size.
Using this method, G91 obtained from the Perseus--Pisces (PP herafter)
redshift survey, a correlation function reproducing the canonical
real--space shape for optical galaxies $\propto r^{-1.8}$ (Davis \&
Peebles 1983), shown in fig.~1.   Although this function is still
in redshift space at large separations, we prefer to use the notation
$\xi(r)$ instead of the conventional $\xi(s)$ to underline the correction
to real space
operated on small scales.  In the same figure we plot (open circles)
the more recent estimate (in pure redshift space) from the extension of the
CfA redshift survey (V92).
Note the the small--scale depression of the latter correlation function
and the remarkable agreement of the two samples for $r>2 \hmpc$.
In particular, both samples are very well described for separations
larger than $r_b\simeq 3.5 \hmpc$ by the simple law $\xi(r) =
[(r/20)^{-0.8} - 1]$ (dashed line), where $r$ is in $\hmpc$.
The consistency of the two surveys further demonstrates that the original
finding of G91 cannot be ascribed to peculiarities in the Perseus--Pisces
region, and that $r_b\simeq 3.5 \hmpc$ seems to represent a
physically meaningful scale (see also Calzetti, Giavalisco \& Meiksin 1992).

By Fourier transforming the above expression for $\xi(r)$ -- assuming
negligible correlations for $s>20\hmpc$ -- we can have a first rough
indication of the shape of the corresponding power spectrum:
$P(k) \propto k^{-2.2}$. Although $\xi(r)$ is not rigorously a power
law at large separations, the corresponding $P(k)$ still seems to be.
More accurately, if we adopt a simple functional form as used by
Peacock (1991) to fit the APM angular function,
$$
P(k)=  {{Ak^{\alpha}}\over{1+ \left({k\over k_c }\right)^{\alpha-n}}}\;\;\;
\eqno(1)
$$
then we can match the observed correlations for $s>r_b$ with $\alpha \simeq
 -2.2$, $k_c \simeq  0.08 \ h \ {\rm Mpc^{-1}}$ and virtually any value of
$n$ in the range $[0,1]$ (which has very little effect on $\xi$ for
$\lambda < 2\pi/k_c$).
At this point we have essentially two ways
to constrain the primordial index $n$.  The first is
simply to assume an Harrison--Zel'dovich primordial spectrum,
i.e. $n=1$. In this case, considering the
microwave background {\sl rms} temperature fluctuation on $10^{\circ}$ scales
measured by COBE, $\sigma_T (10^{\circ} )=[1.085 \pm 0.183]\times 10^{-5}$
(Smoot \etal 1992),
we obtain a normalization corresponding to a {\sl bias factor} $b =
1/\sigma_8 \simeq 1.4$, where $\sigma_8$ is the mass variance in a top--hat
window of $8 \hmpc$ radius.   For clarity, note that we are using here
the conventional definition of $b=\sigma_8(gal)/\sigma_8(mass)$,
together with the Davis \& Peebles (1983) result that gives $\sigma_8(gal)
\simeq 1$.  One may define
a bias factor for bright galaxies, $b^\ast$, compatible with the correlation
functions of fig.~1, considering that, once the redshift space amplification
is corrected using Kaiser (1987) formula, is $\sigma{^\ast}_8(gal)\simeq
1.3$ and therefore $b^\ast \simeq 1.3 b$.

A completely different route to get a plausible value of the $(n,b)$ pair
is to note that if $b=1.8 - 2$, a range which turns out to be
justified in several respects for optical galaxies,
then we have $\sigma_R \sim  1$ for $R\simeq 3.5 \hmpc = r_b$.
In other words, in such a case the
change of slope in the observed correlation function is located right
around the scale where $(\delta \rho / \rho)_{rms}\sim 1$, i.e. coincides
with the expected transition from the linear to the nonlinear clustering
regimes.  This has the important consequence that larger scales should
still be evolving in a quasi--linear fashion and could therefore constrain
directly the shape of the initial spectrum.
Using this kind of bias value, and considering again the COBE constraint
at large wavelengths we obtain $n \simeq 0.75$ if we neglect the
gravitational wave contribution to the Sachs-Wolfe effect (Lucchin, Matarrese
\& Mollerach 1992; A92).  We note that in this case all the important
parameters of the spectrum, i.e. $b\simeq1.9$, $\alpha\simeq -2.2$ and
{ $n\simeq 0.75$ } are globally consistent with those of a correspondingly
tilted CDM spectrum (Cen \etal 1992; A92).  This is
evident in fig.~2a, where we compare our euristic fit with CDM models
of varying primordial index ($n=0.7$, 0.8, 0.9, 1),
adopting the Bardeen \etal (1986, BBKS) CDM transfer function.
It is implicit in this comparison that we are considering our
large--scales fit as representative of the linear power spectrum,
in particular that any residual redshift space effects are
limited to the Kaiser amplification.
This is quite a strong assumption and might be questionable in
several ways (e.g. FDSYH; Bahcall, Cen \& Gramann 1993).  However,
let us assume here as a working hypothesis that expression
(1), with $\alpha=-2.2$, $n=0.75 - 1$, $b=1.4-1.9$, represents
a good linear phenomenological description of the true $P(k)$.
We shall call it the Phenomenologically Induced Model (PIM).
The remainder of the paper will concentrate on comparing it
with the observations, and possibly narrowing the constraints on
its normalization and primordial index.

\bigskip\noindent
{\bf 3. LINEAR AND NONLINEAR TESTS}
\smallskip

The first test is obviously to check whether our phenomenological form
is a good representation
of the `general' behaviour of clustering, as quantified by power spectra
measurements on independent samples.  To this end, in fig.~2b
the PIM, with $n=0.75$ and $n=1$ and normalized to fit the
observed large--scale $\xi(r)$ of the PP sample,
is compared to the direct estimate of $P(k)$ from the CfA survey
(V92).   This comparison is particularly significant, since both surveys
are constructed starting from the same photometric material (the Zwicky
catalogue), and both the PP $\xi(r)$ and the CfA $P(k)$ are estimated on
volume--limited samples with $M$ brighter than $\sim M^\ast$.
For these reasons we prefer to avoid including in the figure measures of
$P(k)$ for, e.g., radio galaxies (Peacock \& Nicholson
1991), or IRAS galaxies (FDSYH; JV93),
which inevitably require an arbitrary amplitude renormalization to
be compared with the optical data (see fig.~1 of V91 for an indirect
comparison).   The agreement of the PIM with the CfA $P(k)$ is
evidently very good.  Note that there is no renormalization between
the curve and the data points.

What seems to be worrying in fig.~2b is that the data points continue
to follow the PIM curve -- which is supposed to be describing the {\sl
linear} $P(k)$ -- also for for $k>0.3\;h\;Mpc^{-1}$, where nonlinear
effects certainly start to become significant.  To understand this,
we have Fourier--transformed the PP $\xi(r)$ of fig.~1, which is
corrected for the main nonlinear distortions produced by the `Fingers
of God'.  The result is plotted as a dashed line in in fig.~2b, showing
that, as one expects from the small--scale $r^{-1.8}$ shape of the correlation
function, $P(k)$ rises to $\sim k^{-1.2}$ as consequence
of nonlinear clustering.  It is interesting to notice, therefore,
that there seems to be a kind of `conspiracy': nonlinear clustering
enhance correlations at small--scales, but when these are viewed in
redshift space, the global effect of small--scale velocity dispersions
seems to bring the spectrum back to a shape similar to the purely linear
one.    This is the same effect visible in the correlation function (fig.~1):
when the `finger of God' effect is not corrected (CfA),
$\xi(r)$ is well approximated by a single law $\propto [(r/20)^{-0.8} -1]$
from small to large scales.

The other important test is to check how the PIM behaves when
nonlinear gravitational evolution sets in.  To this end, we have
performed an N--body simulation, concentrating our attention
on two of the most crucial observational tests: the two--point angular
correlation function $w(\theta)$, and the one--dimensional galaxy
velocity dispersion $\sigma_{1d}$.   We do not perform here a
detailed investigation of redshift--space clustering (as e.g.
in FDSYH), which is left for a more comprehensive paper
involving further simulations with extended dynamic range.

For our specific aim, this first simulation was planned as a compromise
between the necessity of a large box size, required by the
considerable amount of large--scale power present in the PIM spectrum
($L_{box} > 2\pi/k_{turn}\sim 130$ $h^{-1}$ Mpc), and the need of a good
spatial resolution for not to underestimate $\sigma_{1d}$.  We therefore
used a P$^3$M code to integrate the motion of $64^3$ particles within a
$180 \hmpc$ sided
box, with $160^3$ grid points and a resolution of $0.3 \hmpc$.
We used $\Omega_o = 1$ and $H_o = 50\;km\;s^{-1}\;Mpc^{-1}$.
Initial conditions were described by a Gaussian random field, and the
`galaxies' were selected using the BBKS prescription, as in White \etal (1987).

A further important reason for performing an N--body test is to possibly
put some more stringent constraints on $n$ and $b$.  To this end, we
adopted the following procedure. We did not consider the COBE normalization
to fix the present epoch, but leave clustering
evolve until the small--scale shape of $\xi(r)$ for `bright galaxy peaks'
(defined with a threshold corresponding to $\nu_{sim}= 1.3 b \simeq 2.3$)
in the simulation matched the observed one.  At this point
$b= 1/\sigma_8  \simeq 1.8$, and the corresponding $n$ required to
match COBE with the PIM form is $n\simeq0.85$.
Fig.~1 shows the two--point correlation function for the
simulated galaxies (solid line), calculated in real space as
in White \etal 1987.  In order to be compared with the PP
estimate, $\xi(r)$ has been multiplied by the proper linear amplification
factor (Kaiser 1987), on the scales where $(\delta \rho / \rho)_{rms}< 1$.
The match with the observations
is remarkable in both shape and amplitude over a wide range of scales,
a result which seems to support nicely the idea that the change of
slope at $r_b \simeq 3.5 \hmpc$ tags the transition scale to
full nonlinearity.

A crucial test for the model is provided by the angular correlation function
$w(\theta)$, which is by definition independent from any redshift space effect.
We estimated $w(\theta)$  by projecting $\xi(r)$ of the simulation through the
Limber equation, adopting a selection function and normalization corresponding
to the Lick depth (see Peacock 1991).   Since at large separations
($r>15 \hmpc$) the single simulation performed shows slightly less power
than the linear model, on these scales we used $\xi(r)$ from the latter one,
smoothly joined to the numerical result at smaller scales.  The final result
is shown in fig.~3, compared with the
APM (Maddox \etal 1990), and EDSGC (Nichol \& Collins 1993) data.
The PIM spectral shape does still seem to lack some power at very large angular
separations.  This might either suggest that the spatial estimates of $\xi(r)$
and $P(k)$, on which the PIM is based, still suffer from the limited sample
size, or be related to possible sistematics in $w(\theta)$ at these large
angles.

The other main aim of the simulation was to check the small--scale
velocity dispersion produced by the PIM.  We computed
$\sigma_{1d}(r)$ for the dark particles as
$\sigma_{1d}(r)=1/\sqrt{3} \langle | \vvec_1 - \vvec_2 | \rangle ^{1/2}$,
where $\vvec_1$ and $\vvec_2$ are the peculiar velocities of the two
particles with separation $r$.   We obtained $\sigma_{1d}(0.5\hmpc)\simeq
450$ Km s$^{-1}$, rapidly decreasing to $\sim 350$ for $r\ge 2\hmpc$.
This compares interestingly with the
observed $\sigma_{1d}  \simeq 320$ Km s$^{-1}$ on $1 \hmpc$ scales (Fisher
\etal 1993b).

In summary, we have shown that an extremely simplified toy model with
a few specific features, in particular a steep slope $\sim k^{-2.2}$, a
turnover around $k\sim 0.05\; h \;Mpc^{-1}$
and a biasing $b\sim 1.8$, is capable of reproducing remarkably well
some of the main properties of galaxy clustering.   This supports
the soundness of the basic assumptions made for its construction,
in particular the idea that fluctuations on scales larger than $\sim 5
\hmpc$ are still weakly affected by nonlinear evolution.
Clearly, more extended tests are required, in particular direct comparison
of real and redshift space properties from several simulated catalogues.
Also, we have not touched here the issue of which physical scenarios
might be able to produce a spectrum at recombination with this form.
However, it seems that this first phenomenological investigation has
at least underlined a few features which seem to be unavoidable for
any physical model aimed at explaining the observed large--scale
structure.
During the revision of this paper, we became aware of
a work by Baugh \& Efstathiou (1993), who accurately determine
$P(k)$ in true real
space by de--projecting the APM $w(\theta)$.  Their result fully
agrees with our claim that the power spectrum has a steep
range $\sim k^{-2}$ for $0.08<k<0.3\;h\;Mpc^{-1}$ which is not
the result of redshift--space effects, and raises to $k^{-1.2}$
for $k>0.3\;h\;Mpc^{-1}$ due to the onset of nonlinear clustering
(cf. our fig.~2b with, e.g., their fig.~12).

\smallskip
\noindent {\bf {Acknowledgments}}

{\baselineskip 0.5truecm
We thank J.R.~Bond, S.~Bonometto, R.~Carlberg, A.~Klypin, L.~Kofman,
F.~Luc\-chin, J.~Pea\-co\-ck, A.~Provenzale and M.~Scodeggio
for fruitful discussions, P. Haines for reading the
ma\-nu\-scri\-pt, S.~Maddox, R.~Nichol and M.~Vogeley for providing us
with a computer version of their results, CINECA and ICTP
for computing time allocation, and GNA/CNR for financial support.
EB is expecially grateful to P.~Catelan and S.~Matarrese for their
precious suggestions, to SISSA for financial support and to the
Department of Astronomy of the University of Toronto for hospitality
during the final phases of this work.  The referee, M.~Strauss, is
gratefully acknowledged for the many important suggestions which
greatly improved the presentation of this work.
}
\bigskip
\noindent {\bf {References}}

{\baselineskip 0.5truecm
\hi
Adams, F. C., Bond, J. R., Freese, K., Frieman, J. A., \& Olinto, A. V.,
1993, Phys.Rev., D47, 426 (A92).

\hi
Bahcall, N.A., Cen, R., \& Gramann, M., 1993, \apj, 408, L77

\hi
Bardeen, J.M., Bond, J.R., Kaiser, N., \& Szalay, A.S., 1986, \apj, 304, 15.

\hi
Baugh, C.M., \& Efstathiou, G.P., 1993, \mnras, 265, 145.

\hi
Baumgart, D.J., \& Fry, J.N., 1991, \apj, 375, 25.

\hi
Blumenthal, G.R., Faber, S.M., Primack, J.R., \& Rees, M.J., 1984,
\nat, 311, 517.

\hi
Calzetti, D., Giavalisco, M. \& Meiksin, A., 1992, \apj, 398, 429.

\hi
Cen, R., Gnedin, N. Y., Kofman, L. A. \& Ostriker, J. P., 1992, \apj, 399,
L11.

\hi
Cen, R. \& Ostriker, J. P., 1993, preprint.

%
\hi
Davis, M. \& Peebles, P. J. E., 1983, \apj, 267, 456.

%
\hi
Davis, M., Summers, F. S. \& Schlegel, D. 1992, Nature, 359, 393.

\hi
de Lapparent, V., Geller, M. J., \& Huchra J. P., 1988, \apj, 324, 44.

\hi
Efstathiou, G., Sutherland, W. S. \& Maddox, S. J. 1990, Nature, 348, 705.

\hi
Fisher, K. B., Davis, M., Strauss, M. A., Yahil, A. \& Huchra, J. P., 1993,
\apj, 402, 42 (FDSYH).

\hi
Fisher, K. B., Davis, M., Strauss, M. A., Yahil, A. \& Huchra, J. P., 1993b,
\mnras, in press.

\hi
Guzzo, L., Iovino, A., Chincarini, G., Giovanelli, R. \& Haynes, M. P.,
1991, \apj, 382, L5. (G91).

\hi
Hudson, M.J., 1993, \mnras, submitted

\hi
Jing, J. P. \& Valdarnini, R., 1993, \apj, 406, 6 (JV93).

\hi
Kaiser, N., 1987, \mnras, 227, 1.

\hi
Kashlinsky, A., 1992, \apj, 387, L1

\hi
Loveday, J., Peterson, B. A., Efstathiou, G. \& Maddox, S. J., 1992,
\apj, 390,388.

\hi
Lucchin, F., Matarrese, S., \& Mollerach, S., 1992, \apj, 401, L49.

\hi
Maddox, S. J., Efstathiou, G., Sutherland, W. J. \& Loveday, J., 1990,
\mnras, 242, 43p.

\hi
Nichol, R.C., \& Collins, C.A., 1992, \mnras, in press.

\hi
Peacock, J. A., 1991, \mnras, 253, 1P.

\hi
Peacock, J. A. \& Nicholson, D., 1991, \mnras, 253, 307.

\hi
Scaramella, R., 1992, \apj, 390, L57.

\hi
Smoot, G. F. \etal, 1992, \apj, 371, L1.

\hi
Taylor, A. N. \& Rowan-Robinson, M., 1992, \nat, 359, 396.

\hi
Vogeley, M. S., Park, C., Geller, M. J. \& Huchra, J. P., 1992, \apj, 391, L5
(V92).

\hi
White, S.D.M, Frenk, C.S., Davis, M., \& Efstathiou, G., 1987, \apj, 313, 505.
}
\nupa

\noindent {\bf {Figure captions}}
\bigskip
\noindent {\bf {Figure 1.}}

Two--point correlation functions from the PP (G91) and CfA (V92) redshift
surveys.  The CfA estimate is in pure redshift space, while the PP sample
has been statistically corrected for the small--scale `Finger of God'
effect.   Error bars are 1--$\sigma$ bootstrap errors.
The dashed line shows the curve $\xi(r) = (r/20)^{-0.8}-1$,
while the solid line is the result of the P$^3$M N-body integration
of the PIM.

\smallskip
\noindent {\bf {Figure 2.}}

a) The power spectrum of the phenomenologically induced model (PIM), with
primordial index fixed to $n=0.75$, normalized to COBE and compared to flat
CDM models with varying primordial index ($n=0.7$, 0.8, 0.9, 1).

b) The PIM power spectrum normalized to match the PP galaxy correlations of
Fig.~1 (solid line) for the two cases $n=0.75$ and $n=1$, compared to the
direct estimate
of $P(k)$ from the CfA survey by V92 (circles).   The dashed line shows the
result of Fourier transforming the observed PP small--scale $\xi(r)$.

\smallskip
\noindent {\bf {Figure 3.}}

Angular correlation function $w(\theta)$ calculated from the N--body
integration of the PIM (solid line), compared with
the APM (dots) and EDSGC (circles) data.   The EDSGC 1$\sigma$ error bars
show the large uncertainty existing in $w(\theta)$ for $\theta > 10^{\circ}$.

\nupa
\noindent{\bf Postal and e-Mail Addresses}
\bigskip
{\baselineskip 0.5truecm
\noindent

\noindent Enzo Branchini\lb
SISSA, Strada Costiera 11,\lb
I-34014 TRIESTE - ITALY\lb
e-mail: enzo@tsmi19.sissa.it
\bigskip

\noindent Luigi Guzzo\lb
Osservatorio Astronomico di Brera, Sezione di Merate \lb
Via Bianchi 46, I-22055 MERATE (CO) - ITALY\lb
e-mail: guzzo@astmim.mi.astro.it
\bigskip

\noindent Riccardo Valdarnini\lb
SISSA, Strada Costiera 11,\lb
I-34014 TRIESTE - ITALY\lb
e-mail: valda@tsmi19.sissa.it

}

\end